\documentclass[10pt,aps,pre,twocolumn,superscriptaddress,showpacs,showkeys]{revtex4-2}
\usepackage{color}
\usepackage{graphicx}
\usepackage{hyperref}
\usepackage{amsmath,amssymb}
\usepackage{epstopdf}
\usepackage{subcaption}
\graphicspath{{figs/}}
\usepackage{float}
\usepackage[singlelinecheck=false,justification=raggedright]{caption}
\usepackage{dcolumn}   % needed for some tables
\usepackage{bm}        % for math
\usepackage{verbatim}

\begin{document}
\title{Depinning of KPZ Interfaces in Fractional Brownian Landscapes}

\author{N. Valizadeh}
%\email{valizadeh.neda1204@gmail.com}
\affiliation{Department of Physics, University of Mohaghegh Ardabili, P.O. Box 179, Ardabil, Iran}
\author{M. N. Najafi}
\email{morteza.nattagh@gmail.com}
\affiliation{Department of Physics, University of Mohaghegh Ardabili, P.O. Box 179, Ardabil, Iran}

\begin{abstract}
We explore the critical dynamics of driven interfaces propagating through a two-dimensional disordered medium with long-range spatial correlations, modeled using fractional Brownian motion (FBM). Departing from conventional models with uncorrelated disorder, we introduce quenched noise fields characterized by a tunable Hurst exponent \( H \), allowing systematic control over the spatial structure of the background medium. The interface evolution is governed by a quenched Kardar–Parisi–Zhang (QKPZ) equation modified to account for correlated disorder, namely QKPZ$_H$. Through analytical scaling analysis, we uncover how the presence of long-range correlations reshapes the depinning transition, alters the critical force \( F_c \), and gives rise to a family of critical exponents that depend continuously on \( H \). Our findings reveal a rich interplay between disorder correlations and the non-linearity term in QKPZ$_H$, leading to a breakdown of conventional universality and the emergence of nontrivial scaling behaviors. The exponents are found to change by $H$ in the anticorrelation regime ($H<0.5$), while they are nearly constant in the correlation regime ($H>0.5$), suggesting a super-universal behavior for the latter. By a comparison with the quenched Edwards-Wilkinson model, we study the effect of the non-linearity term in the QKPZ$_H$ model. This work provides new insights into the physics of driven systems in complex environments and paves the way for understanding transport in correlated disordered media.\\

Keywords: Depinning transition, Fractional Brownian motion, Long-range correlations, Interface dynamics, Critical exponents, Non-equilibrium statistical physics
\end{abstract}

%\pacs{05., 05.20.-y, 05.10.Ln, 05.45.Df}
%\keywords{{\color{blue}Depinning Transition, porous media}, fluid dynamics, critical exponents}

\maketitle

\section{Introduction}

The dynamics of driven interfaces in disordered environments constitute a rich area of research within non-equilibrium statistical physics. Such systems are not only of theoretical interest but also manifest in a wide range of interdisciplinary applications. Examples span diverse fields, including fluid displacement in porous media~\cite{rubio1989self, horvath1991anomalous, primkulov2019signatures, armstrong2015modeling, hashemi1999dynamics, rabbani2017new}, flame propagation~\cite{zhang1992modeling}, bacterial colony growth~\cite{moglia2016pinning}, forest fires~\cite{cheraghalizadeh2025forest} and magnetic flux line motion in type-II superconductors~\cite{blatter1994vortices}. Beyond natural phenomena, these ideas are foundational in emerging technologies such as nano-patterning, microfluidics, and the development of functional surfaces and materials~\cite{pismen2006patterns, collet2014instabilities, cross1993pattern}.

A fundamental feature in these systems is the \textit{depinning transition}, a critical point at which an interface, initially stuck due to the presence of quenched disorder, begins to move under the influence of a driving force. Below the critical force $F_c$, the interface remains pinned, and above it, it propagates with a non-zero average velocity. This transition exhibits universal behavior characterized by scale invariance, critical exponents, and diverging correlation lengths, making it a canonical example of a non-equilibrium phase transition~\cite{rosso2001monte, duemmer2005critical, rosso2002roughness, nattermann1992dynamics, leschhorn1997driven, kolton2005nonequilibrium}.

Several universality classes have been identified for such transitions, each associated with distinct scaling laws and microscopic dynamics. The \textit{quenched Edwards--Wilkinson} (QEW) class~\cite{valizadeh2023edwards,valizadeh2021edwards,ferrero2013numerical, le2002two} describes interfaces with purely diffusive dynamics, while the \textit{quenched Kardar--Parisi--Zhang} (QKPZ) class~\cite{rosso2003depinning, ferrero2020creep} includes the effects of local growth anisotropy. The \textit{directed percolation depinning} (DPD) class~\cite{leschhorn1996anisotropic, amaral1995scaling} captures situations where pinning paths dominate the large-scale structure. These classes are typically distinguished by measuring critical exponents such as the roughness, growth, and dynamic exponents, associated with the roughness of the interface~\cite{amaral1995scaling,yang1993diffraction,patrikar2004modeling}. The velocity exponent, is another exponent which governs the spatiotemporal evolution of the interface~\cite{ferrero2013nonsteady,mukerjee2023depinning}.

Despite the sophistication of existing models, most assume that the underlying disorder is \textit{uncorrelated} or possesses only short-range correlations (see~\cite{amaral1995scaling} for a good review). In practice, however, many disordered media---especially natural ones like geological formations, porous rocks, and biological tissues---exhibit complex internal structures with long-range spatial correlations~\cite{makse1996long,sarma2015using,wang2020anomalous,kolton2018critical,zierenberg2017percolation,rohfritsch2021propagation}. Properties such as porosity, permeability, or fracture distribution often follow \textit{scale-invariant or fractal patterns}, which cannot be captured by simple uncorrelated noise. The influence of such \textit{long-range correlated quenched disorder} on the depinning transition remains an open and pressing question.

Some theoretical progress has been made in this direction. Using \textit{dynamical renormalization group} techniques, it has been shown that long-range correlations in disorder can fundamentally alter the universality class of the depinning transition~\cite{fedorenko2006statics}. In parallel, numerical and empirical studies have examined how correlated structures affect anomalous transport properties~\cite{wang2020anomalous}, percolation thresholds~\cite{zierenberg2017percolation}, and interface morphology~\cite{kolton2018critical}. Nevertheless, a comprehensive understanding of how spatial correlations in the host medium modify depinning behavior is still lacking.

To incorporate correlated disorder into models of interface dynamics, several approaches have been developed. These include long-range correlated percolation lattices~\cite{najafi2018coupling, najafi2016monte, najafi2016water}, Ising-like disordered media~\cite{cheraghalizadeh2018self}, and Gaussian random fields with Coulomb-like correlations~\cite{valizadeh2021edwards, cheraghalizadeh2018gaussian1}. Among these, \textit{fractional Brownian motion} (FBM) has emerged as a powerful and versatile tool. FBM generates random fields with tunable long-range correlations controlled by the \textit{Hurst exponent} $H$, where $H > 0.5$ corresponds to positive correlations (smoother surfaces), $H = 0.5$ recovers standard Brownian motion (uncorrelated), and $H < 0.5$ leads to negative correlations (rougher surfaces)\cite{cheraghalizadeh2024fractional}. FBM has been successfully applied to model porosity and permeability distributions in geological reservoirs~\cite{sahimi1996scaling, hamzehpour2006generation}, traffic in networks, anomalous diffusion, and financial time series~\cite{leland1994statistical,biagini2008stochastic,pacheco2024fractional,house2025fractional,mandelbrot1968fractional}.

In this work, we focus on the QKPZ interface dynamics in a two-dimensional disordered medium where the quenched disorder is generated using FBM, namely QKPZ$_H$. By systematically varying the Hurst exponent $H$, we examine how the strength and nature of correlations in the host medium influence the critical force $F_c$, the temporal and spatial scaling behavior of the interface, and the corresponding universality class. Our goal is to explore whether correlated quenched disorder drives the system away from the standard QKPZ universality and toward a new scaling regime. Our results reveal nontrivial dependencies of the depinning threshold and scaling exponents on the correlation strength and demonstrate that long-range disorder fundamentally reshape the critical dynamics of driven interfaces. This study contributes to a broader understanding of how realistic structural features in complex systems influence transport, growth, and dynamical phase transitions. It also opens up new directions for modeling natural systems where disorder is not only present but structured in a scale-invariant way.

The paper structure is as follows: in the next section we introduce the general properties of driven interfaces and the depinning transition. Section~\ref{sec:FBM} is devoted to the construction of correlated random media via fractional Brownian motion (FBM) and its statistical properties. The scaling behavior of the interface velocity and the anomalous roughness dynamics are analyzed in Section~\ref{sec:Scaling} and ~\ref{sec: anomalous roughness} , where we consider both numerical and analytical results across different correlation regimes. Section~\ref{sec:Comparison} presents a detailed comparison between the QKPZ$_H$ and QEW$_H$ models, as well as other universality classes, highlighting the role of disorder correlations and nonlinearity. The paper is closed by concluding remarks in Section~\ref{sec:Conclusion}.

\section{General Properties of Driven Interfaces}\label{SEC:General}

When a fluid starts to flow through a porous medium under the influence of a driving force, its motion is governed by the competition between the fluid dynamics and the inherent disorder of the medium. If the driving force is below the critical threshold, the fluid moves initially but eventually becomes pinned by the obstacles present in the environment. On the other hand, if the driving force is much larger than the critical threshold, the system enters a moving phase. The point that separates these two regimes, from zero terminal velocity to the onset of nonzero velocity is referred to as the depinning transition.\\ 
The depinning transition is an important concept in statistical physics, particularly in the study of non-equilibrium critical phenomena, and has long been the subject of extensive theoretical and experimental research,  such as immiscible fluid invasion in porous substrates~\cite{stokes1988interface,rubio1989self,horvath1991dynamic} and capillary rise in fibrous materials like paper~\cite{buldyrev1992anomalous,amaral1994new,amaral1995avalanches}. These studies report widely varying critical exponents and highlight rich dynamical features, including $1/f$ noise~\cite{krug19911} and anomalous fluctuations~\cite{horvath1991anomalous}. This concept states that when the external driving force applied to a fluid or any mobile object in a disordered medium exceeds a certain critical threshold, the system begins to move. The depinning transition characterizes the scaling properties of the system at the onset of motion, which can reveal the statistical features of the underlying dynamics. \\
The study of kinetic roughening and surface growth processes has led to the identification of distinct universality classes, each characterized by scaling exponents and symmetry properties. Two paradigmatic stochastic growth models are the EW and  KPZ equations, which describe the time evolution of a height field  representing a growing interface. Their quenched counterparts (QEW and QKPZ) extend these models to disordered media with static, spatially inhomogeneous pinning forces, capturing the physics of depinning transitions. We focus on the QKPZ model that the interface dynamics are governed by the quenched KPZ equation:
\begin{equation}
\frac{\partial h}{\partial t} = F + \nu \nabla^2 h + \frac{\lambda}{2} |\nabla h|^2 + \eta(x,h),
\label{QKPZ1}
\end{equation}
where $F$ is a constant driving force , $\nu$ is the surface tension coefficient, $\eta(x,h)$ is quenched disorder and  $\lambda$ controls the strength of the nonlinearity. \\
At criticality, the interface becomes self-affine. A central observable is the interface width (or roughness), defined as:
\begin{equation}
	w^2(T, F, L) = \left\langle \overline{(h(x) - \bar{h})^2} \right\rangle,
\end{equation}
where $\bar{h}$ is the spatially averaged height. For any value of $F$, the evolution of $w$ exhibits two distinct regimes: a growth regime for short times $T \ll T_X$ with $w \sim T^{\beta_w}$, and a saturation regime for long times $T \gg T_X$ with $w \sim L^{\alpha_w}$. The crossover time scales as $T_X \sim \xi_F^{z_w}$, where $\xi_F$ is the correlation length and $z_w = \alpha_w/\beta_w$ is the dynamic exponent.

The finite-size scaling form is given by:
\begin{equation}
	W(T, F, L) = L^{\xi}\mathcal{P}\left(\frac{T}{L^{z}}\right),
    \label{Eq:roughness}
\end{equation}
with the universal function $\mathcal{P}(x)$ obeying:
\begin{equation}
	\mathcal{P}(x) \sim
	\begin{cases}
		x^{\xi/z}, & y \ll 1, \\
		\text{const.}, & y \gg 1.
	\end{cases}
\end{equation}

At the depinning threshold ($F = F_c$), the average velocity decays with time as a power law:
\begin{equation}
	v_c(T, L \rightarrow \infty) \sim T^{-q}.
    \label{Eq:Power-Law}
\end{equation}
Below threshold ($F < F_c$), the interface velocity decays exponentially:
\begin{equation}
	v(T, F, L \rightarrow \infty) \sim e^{-T / T_X} = e^{-T / \zeta_F^{z_w}}.
\end{equation}
where $\zeta_F$ is the correlation length which diverges as the system approaches criticality:
\begin{equation}
	\zeta_F \sim |F - F_c|^{-\nu},
\end{equation}
and for $F > F_c$, the steady-state velocity scales as:
\begin{equation}
	v_\infty(F, L \rightarrow \infty) \sim f^{\mu}, \ f = \frac{F - F_c}{F_c},
     \label{Eq:criticalVelocity}
\end{equation}
where $\mu$ is the velocity exponent. Assuming that interface motion results from correlated unpinning events of scale $\xi_F$, the hyperscaling relation follows:
\begin{equation}
	v_\infty \sim \frac{w_\text{sat}}{T_X} \sim \frac{\zeta_F^{\xi}}{\zeta_F^{z}} \sim f^{\nu(z - \xi)},
\end{equation}
yielding:
\begin{equation}
\mu = \nu(z - \xi).
\end{equation}

Accurate determination of the critical exponents and verification of hyperscaling relations is essential for identifying the underlying universality class.\\
Although significant progress has been made in understanding depinning phenomena, the role of the host medium has received comparatively less attention. A key factor in these systems is the nature of noise correlations, which can vary across different physical settings. Long-range correlations have been analyzed in the depinning transition of elastic manifolds embedded in disordered hosts using dynamical renormalization group techniques~\cite{fedorenko2006statics}. Such correlations are also commonly observed in porous media, influencing properties such as porosity~\cite{kolton2018critical,adler2013fractures,hardy1994fractals}, diffusion~\cite{wang2020anomalous}, and permeability~\cite{zierenberg2017percolation,sahimi1993flow,najafi2018coupling,najafi2016monte}.  

Several approaches exist for incorporating long-range correlated noise into random media as a basis for dynamical modeling. Examples include percolation-based models~\cite{najafi2016water,cheraghalizadeh2018self,cheraghalizadeh2018gaussian}, and Gaussian random Coulomb potentials~\cite{valizadeh2021edwards,cheraghalizadeh2018gaussian}, where correlations often exhibit power-law behavior in certain regions of phase space. Another important framework is provided by FBM~\cite{cheraghalizadeh2024fractional}, which has been used to represent quenched long-range disorder.\\

The aim of this paper is to systematically examine how correlations in disordered host media influence the depinning transition. To this end, the host medium is modeled using FBM, where the correlations are characterized by the Hurst exponent, which governs the roughness and associated statistical properties.

\section{Correlated Random Media via Fractional Brownian Motion}\label{sec:FBM}

In this study, we investigate the dynamics of a driven interface propagating through a random host medium. Unlike previous studies that typically assume uncorrelated (white) disorder, we consider a \textit{correlated} quenched random landscape, modeled using two-dimensional fractional Brownian motion (2D FBM). Before introducing the dynamical model, we first provide a brief overview of FBM and explain how it is used to construct the correlated disorder field.

Our model considers interface motion in a $1+1$ dimensional geometry (i.e., scalar displacement over a two-dimensional substrate) where the underlying disorder exhibits long-range spatial correlations. These correlations are characterized by the Hurst exponent $H$ and generated via FBM. The dynamical evolution of the interface is then modeled by the QKPZ$_H$ equation, which will be detailed in the following section.

\begin{figure*}[t]
    \centering
    \begin{subfigure}{0.45\textwidth}
        \includegraphics[width=\textwidth]{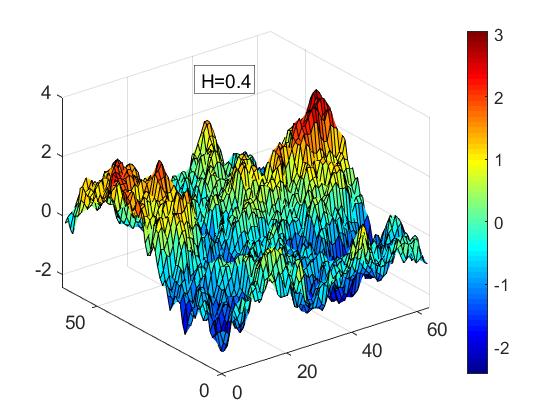}
      %  \caption{$H = 0.4$}
        \label{fig:FBM_H04}
    \end{subfigure}
    \hfill
    \begin{subfigure}{0.45\textwidth}
        \includegraphics[width=\textwidth]{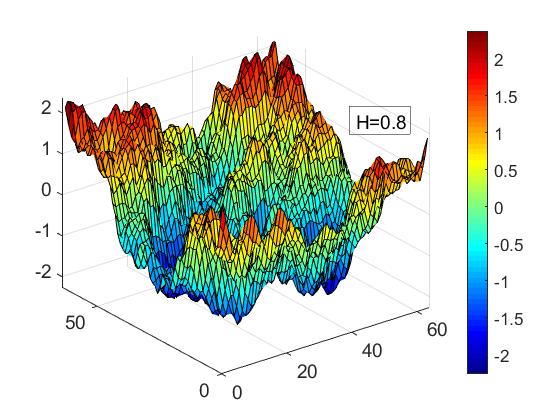}
      %  \caption{$H = 0.8$}
        \label{fig:FBM_H08}
    \end{subfigure}
    \caption{Examples of 2D FBM surfaces with different Hurst exponents. Left: rougher landscape with anti-correlated noise ($H = 0.4$). Right: smoother surface with positively correlated increments ($H = 0.8$).}
    \label{fig:FBM}
\end{figure*}

The Hurst exponent $H$ controls the nature of the correlations: for $H=0.5$, the increments of FBM are uncorrelated (standard Brownian motion), while $H>0.5$ ($H<0.5$) corresponds to positively (negatively) correlated disorder, resulting in smoother (rougher) surfaces, as illustrated in Fig.~\ref{fig:FBM}.

 FBM is a Gaussian stochastic process with stationary, power-law correlated increments. It is widely used as a prototypical model for anomalous diffusion and correlated random media, with applications ranging from polymer dynamics~\cite{chakravarti1997fractional,panja2010generalized} and intracellular transport~\cite{szymanski2009elucidating} to traffic in electronic networks~\cite{mikosch2002network} and financial time series~\cite{rostek2013note}.

The 2D FBM field $B_H(\mathbf{r})$, where $\mathbf{r} = (x, y)$, is defined such that
\begin{equation}
    \langle B_H(\mathbf{r}) - B_H(\mathbf{r}_0) \rangle = 0, \qquad
    \langle [B_H(\mathbf{r}) - B_H(\mathbf{r}_0)]^2 \rangle = |\mathbf{r} - \mathbf{r}_0|^{2H},
    \label{Eq:def}
\end{equation}
where $H$ is the Hurst exponent controlling the degree of correlation.

To numerically generate 2D FBM fields on an $L_x \times L_y$ lattice, we use the fast Fourier transform method. The approach begins by generating a Gaussian white noise field $\zeta(\mathbf{r})$ with the following statistical properties:
\begin{equation}
    \langle \zeta(\mathbf{r}) \rangle = 0, \quad
    \langle \zeta(\mathbf{r}) \zeta(\mathbf{r}') \rangle = D \, \delta^2(\mathbf{r} - \mathbf{r}'),
    \label{Eq:zeta}
\end{equation}
where $D$ controls the disorder strength.

We then define the Fourier-transformed FBM field as:
\begin{equation}
    \tilde{B}_H(\mathbf{k}) = \left(k_x^2 + k_y^2 \right)^{-(H+1)/2} \tilde{\zeta}(\mathbf{k}),
    \label{Eq:fBM}
\end{equation}
with
\begin{equation}
    \tilde{\zeta}(\mathbf{k}) = \frac{1}{\sqrt{L_x L_y}} \sum_{\mathbf{r}} \zeta(\mathbf{r}) e^{i \mathbf{k} \cdot \mathbf{r}}.
\end{equation}

The real-space FBM field is then obtained via the inverse FFT:
\begin{equation}
    B_H(\mathbf{r}) = \frac{1}{\sqrt{L_x L_y}} \sum_{\mathbf{k}} \tilde{B}_H(\mathbf{k}) e^{-i \mathbf{k} \cdot \mathbf{r}}.
    \label{Eq:fBM-Def}
\end{equation}

Importantly, this construction satisfies the self-affine scaling property:
\begin{equation}
    B_H(\alpha \mathbf{r}) \overset{d}{=} \alpha^H B_H(\mathbf{r}),
    \label{Eq:scaling}
\end{equation}
where $\overset{d}{=}$ denotes equality in distribution. The scaling behavior can be derived from the transformation properties of the white noise field $\zeta(\mathbf{r})$ under spatial rescaling (see Appendix \ref{App:Scaling})
\begin{equation}
    \zeta(\alpha \mathbf{r}) \overset{d}{=} \alpha^{-1} \zeta(\mathbf{r}), \quad
    \tilde{\zeta}(\alpha^{-1} \mathbf{k}) \overset{d}{=} \tilde{\zeta}(\mathbf{k}),
\end{equation}
which, together with Eq.~\ref{Eq:fBM}, implies:
\begin{equation}
    \tilde{B}_H(\alpha^{-1} \mathbf{k}) \overset{d}{=} \alpha^{H+1} \tilde{B}_H(\mathbf{k}),
\end{equation}
leading to the scaling form in Eq.~\ref{Eq:scaling}.

The power spectral density of the resulting FBM landscape in 2D is given by:
\begin{equation}
    S(\mathbf{k}) = \frac{a}{\left(k_x^2 + k_y^2\right)^{(H+1)/2}},
    \label{Eq:fbm}
\end{equation}
where $a$ is a normalization constant.\\

In this paper we generate a correlated energy landscape using a spectral method for two-dimensional FBM and evolve a one-dimensional interface under the influence of both a constant driving force and the FBM-generated noise. We model the system using a one-dimensional interface $h(x,T)$ evolving in time $T$ over a two-dimensional substrate. The spatial coordinate is $\mathbf{r} = (x,y)$, and the interface initially lies flat at $y = 0$. The medium’s structural disorder is represented by a 2D FBM field $B_H(\mathbf{r})$, over which the QKPZ equation governs the motion of a driven interface. We analyze the roughness, velocity, and scaling behavior across a range of $H$ values. identify critical behavior through data collapse, scaling functions, and exponent extraction. This approach allows us to explore how the strength and nature of spatial correlations in the disorder—encoded via the Hurst exponent—affect interface dynamics, depinning transitions, and universality classes.

We call the QKPZ on top of the FBM as QKPZ$_H$. The interplay between the nonlinear growth term in the QKPZ equation and the correlated disorder landscape leads to rich physics, which we explore in the following sections.
\begin{comment}
For instance, the QEW class (with uncorrelated noise) exhibits $\alpha \approx 1.25$ and $\theta \approx 0.25$~\cite{kim2006depinning,lopez1997interface,makse1995scaling}. Differentiating between QEW and QKPZ universality classes may also be achieved by analyzing the system's response to tilted initial conditions~\cite{amaral1994universality,moglia2014interfacial}.
\end{comment}

\begin{figure*}[!htbp]
	\begin{subfigure}{0.32\textwidth}\includegraphics[width=\textwidth]{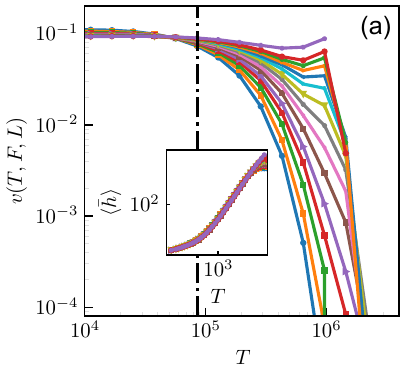}
	\end{subfigure}
	\begin{subfigure}{0.32\textwidth}\includegraphics[width=\textwidth]{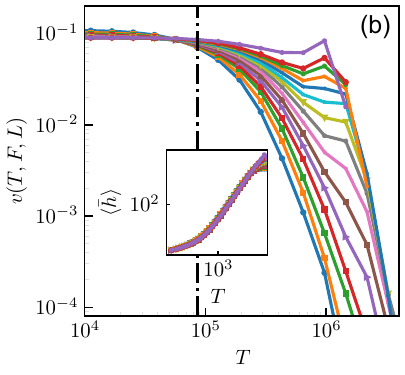}
	\end{subfigure}
	\begin{subfigure}{0.32\textwidth}\includegraphics[width=\textwidth]{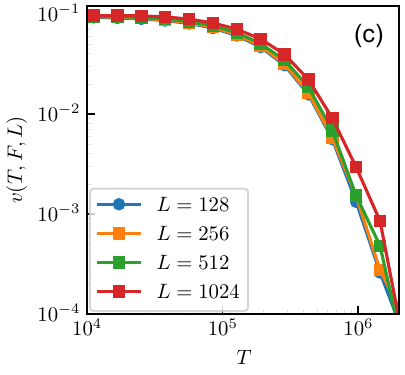}
	\end{subfigure}
	%\begin{subfigure}{0.45\textwidth}\includegraphics[width=\textwidth]{f11_h8.pdf}
	%\end{subfigure}
	\caption{ Log-Log plots of velocity $v(T,L,F)$ as function of time $T$ as obtained for different driving forces $F$ (from down to up $F=0.7$,  $0.8$, ,..., $1.9$, $2$, $2.5$) that measured for a system size $L = 1024$.  After a short time, for $F>F_{c}$ The interface starts moving and for  $F<F_{c}$ the interfaces become pinned. For all figures, the results are obtained by starting with flat interfaces and averaged for 20000 realizations. In lower insets are log-log plots of the average height $\left\langle\bar{h}\right\rangle $ versus time $T$ for various rates of the driving force. In fact, the velocity strongly depends on the driving force $F$. (a) For $ H=0.4$ and (b) $H=0.8$. Log-Log plots of $v(T,L,F)$ versus $T$ at the driving force $F=F_c$, as measured for different system sizes $L$ for (c) $H=0.4$, $F=1.2$. The dashed line corresponds to the crossover point $T^*$, which is further confirmed by the scaling analysis in Fig.~\ref{Fig:vstar}.
}
	\label{fig:haveT}
\end{figure*}
\begin{figure*}[!htbp]
	\begin{subfigure}{0.4\textwidth}\includegraphics[width=\textwidth]{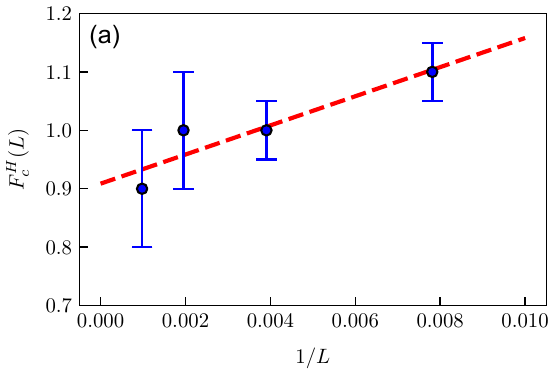}
	\end{subfigure}
	\begin{subfigure}{0.4\textwidth}\includegraphics[width=\textwidth]{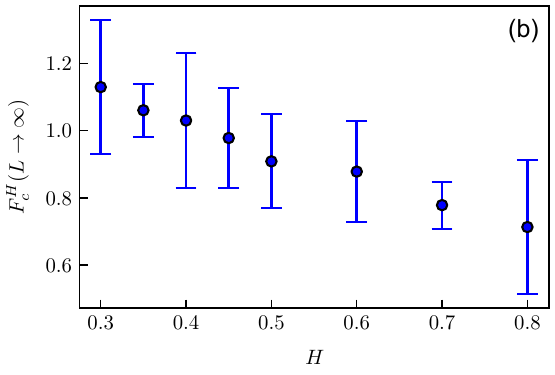}
	\end{subfigure}
	%\begin{subfigure}{0.45\textwidth}\includegraphics[width=\textwidth]{thetaf.pdf}
	%\end{subfigure}
	\caption{(a): Finite-size scaling analysis for $H=0.5$. The plot shows the extrapolation of critical force $F_c(L)$ versus inverse system size $1/L$. The linear fit (dashed line) yields the thermodynamic limit estimatethrough the intercept.(b): Dependence of critical force $F_c$ on Hurst exponent $H$. }
	\label{Fc}
	%\end{center}
\end{figure*}

\begin{figure}[!htbp]
    \centering
    \includegraphics[width=0.45\textwidth]{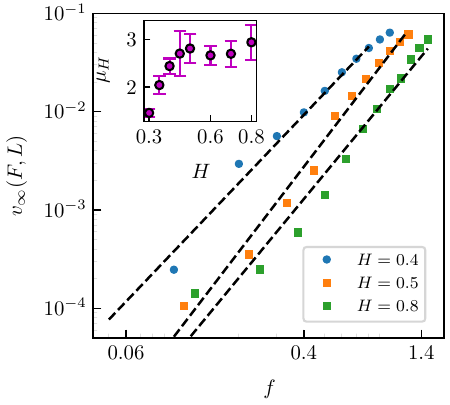}
    \caption{Steady-state velocity $v_\infty$ as a function of reduced force $f=(F-F_c)/F_c$ on a log–log scale for different values of the Hurst exponent $H$ for $L=1024$. The extracted velocity exponent $\mu_H$ for $L\to\infty$.}

    \label{fig:vinf}
\end{figure}
\begin{figure*}[!htbp]
	\begin{subfigure}{0.32\textwidth}\includegraphics[width=\textwidth]{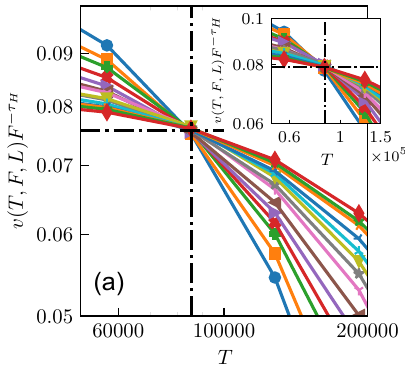}
	\end{subfigure}
	\begin{subfigure}{0.32\textwidth}\includegraphics[width=\textwidth]{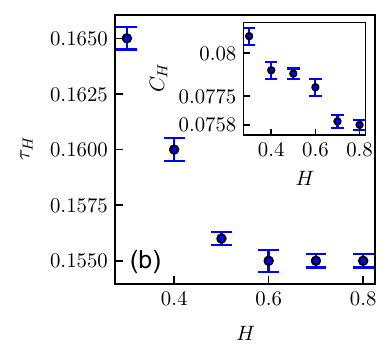}
	\end{subfigure}
	\begin{subfigure}{0.32\textwidth}\includegraphics[width=\textwidth]{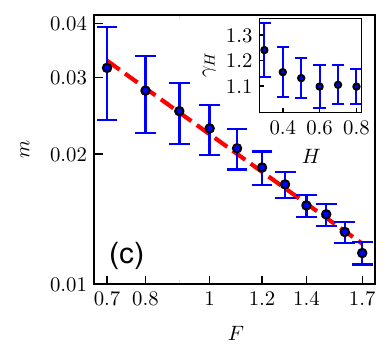}
	\end{subfigure}
	
	\caption{Scaling of the velocity near criticality. (a) Data collapse of the rescaled velocity $v F^{-\tau_H}$ versus time $T$ for $H=0.8$ and $H=0.4$ that is in inset  , showing a universal curve beyond the crossover time $T^\ast$. (b) --Scaling exponents $\tau_H$, and $C_H$ on the Hurst exponent $H$. (c) the slope $m$ of part (a) around $T=T^*$, which is expected to be $\frac{B}{F^{\gamma_H}}$ for $H=0.8$ according to Eq.~\ref{eq:v_scaling}. Inset shows $\gamma_H$ in terms of $H$.}
	\label{Fig:vstar}
	%\end{center}
\end{figure*}
\begin{figure*}[!htbp]
	\begin{subfigure}{0.25\textwidth}\includegraphics[width=\textwidth]{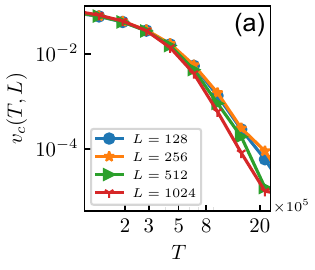}
	\end{subfigure}
	\begin{subfigure}{0.25\textwidth}\includegraphics[width=\textwidth]{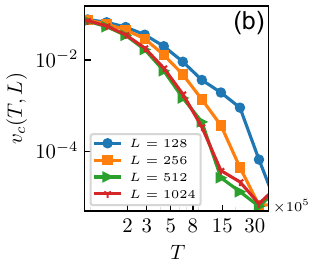}
	\end{subfigure}
	\begin{subfigure}{0.24\textwidth}\includegraphics[width=\textwidth]{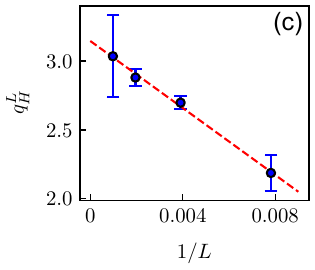}
	\end{subfigure} \begin{subfigure}{0.24\textwidth}\includegraphics[width=\textwidth]{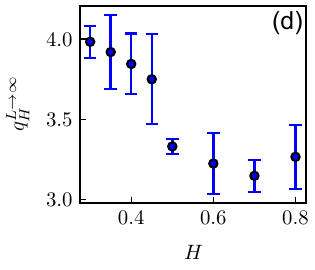}
	\end{subfigure}
	%\begin{subfigure}{0.45\textwidth}\includegraphics[width=\textwidth]{thetaf.pdf}
	%\end{subfigure}
	\caption{Scaling of the critical velocity  $v_c(T, L)$ at the depinning threshold.
        (\textbf{a}, \textbf{b}) Log-log plots of the critical velocity versus time $T$ at $F = F_c$ for (\textbf{a}) $H = 0.4$ and (\textbf{b}) $H = 0.8$, showing power-law decay $v_c \sim T^{-q_H}$.
        (\textbf{c}) Finite-size scaling analysis for $H=0.7$, (\textbf{d})  Dependence of the decay exponent $q_H$ on the Hurst exponent $H$, indicating slower relaxation (a smaller $q_H$) for more strongly correlated disorder.
      }
	\label{Fig:vc}
	%\end{center}
\end{figure*}
\section{Scaling Relation for the Velocity of QKPZ$_H$.}\label{sec:Scaling}

To characterize the dynamical behavior of the driven interfaces in the presence of correlated quenched disorder, QKPZ$_H$, we analyze the time evolution of the interface velocity. This section presents a detailed investigation of how the velocity scales with the driving force $F$, system size $L$, and the Hurst exponent $H$.\\

The general behavior of $v$ in terms of $T$ and $F$ for a fixed $L=1024$ is shown in Figs.~\ref{fig:haveT}a ($H=0.4$) and~\ref{fig:haveT}b ($H=0.8$), while the $L$-dependence is represented in Fig.~\ref{fig:haveT}c. Insets show the mean height $\langle \bar{h} \rangle$ in terms of time. A key observation in our simulations is the existence of a characteristic \textit{crossover time} $T^{*}$, separating two distinct dynamical regimes. $T^{*}$, as represented by dashed lines in Figs.~\ref{fig:haveT}a and~\ref{fig:haveT}b, is mathematically defined as the point where the behavior of $v$ in terms of $F$ changes in nature: before (after) $T^*$ the velocity increases (decreases) with increasing $F$ at a fixed $T$ and $L$. This is the case for all values of $H$. More precisely, we found that the velocity in a piecewise form:

\begin{equation}
v_{H}(T, F, L) =
\begin{cases}
v_< (T, F) & T < T^* \\
v^*(F) & T = T^* \\
v_> (T, F, L) & T > T^*
\end{cases}
\label{eq:velocity_piecewise}
\end{equation}
where $v^*$ is a $L$-independent velocity that is different from $v_<{(T,F)}$ in the sense that it shows power-law behavior in terms of $F$, which is analyzed later. A scaling analysis gives us both $T^*$ and $v^*$, ending up with some $H$-dependent exponent (following sections). Note also that for $T < T^*$ the dependence of $v_<$ on $L$ is negligible, while for $T > T^*$, the velocity shows dependence on $L$.\\

The fate of the interfaces, i.e. their asymptotic velocity in terms of $T$ is determined by $F$: there is an $H$-dependent critical force $F^H_c$ so that for $F<F^H_c$, the velocity decays to zero representing the pinned phase, while for $F>F^H_c$ it saturates at a finite value (moving phase). Panel (c) demonstrates the velocity in terms of $T$ at criticality ($F=F^H_c$), revealing power-law decay characteristic of scale-invariant dynamics in the depinning transition point (see Eq.~\ref{Eq:Power-Law}).\\

To determine the critical force $F_c$, we employ the \emph{velocity ratio method}, which is computationally efficient. This method is based on the steady-state velocity $v_\infty(F,L) = \lim_{T \to \infty} v_{H}(T, F, L)$. For this end, we determine the velocity of two successive  driving forces $F_1 < F_2$, and define the logarithmic velocity ratio as:
\begin{equation}
x(L, F_1) \equiv \ln \left( \frac{v_\infty(F_2,L)}{v_\infty(F_1,L)} \right).
\end{equation}
The critical force $F^H_c(L)$ is then extracted via the threshold condition:
\begin{equation}
F^H_c(L) = \sup \left\{ F_1 \mid x(L, F_1) \geq x_{\text{th}} \right\},
\end{equation}
where we empirically set $x_{\text{th}} = 1$.  Then the thermodynamic value for $F_c$ is obtained using the extrapolation
\begin{equation}
F^H_c=\lim_{L\to \infty}F^H_c(L).
\end{equation}
We found this systematic method appropriate for the cases where, due to the fluctuations it is challenging to extract the critical point using scaling arguments.\\

In Fig.~\ref{Fc}a we plot $F_c(L)$ in terms of $1/L$ for $H=0.5$, so that the $F_c(L\to\infty)$ is obtained as a fitting parameter:
\begin{equation}
F^H_c(L)=F^H_c(L\to\infty)+\frac{A_H}{L},
\end{equation}
where $A_H$ is some $L$-independent non-universal fitting parameter, that is not important in our analysis. Figure~\ref{Fc}b displays $F^H_c$ as a function of $H$, revealing a clear monotonic decrease. This trend can be readily understood by noting that stronger disorder correlations demand a larger driving force to overcome pinning. This decrease is found to be according to the following linear relation
\begin{equation}
F^H_c(L\to\infty)=F_c^{H=0}-\left[F_c^{H=0}-F_c^{H=1}\right]H
\label{Eq:Fc}
\end{equation}
where $F_c^{H=0}=1.35\pm 0.05$ and $F_c^{H=1}=0.54\pm 0.05$.\\

Near the transition point $F\sim F_c$, the system exhibits scale invariance. The interface velocity, particularly in the thermodynamic limit ($L \to \infty$), follows the power-law relation Eq.~\ref{Eq:criticalVelocity} according to which
\begin{equation}
v_\infty(F,L) \sim (F - F_c)^{\mu_{H}},
\label{eq:v_crit}
\end{equation}
where $\mu^{}_{H}$ is an $H$-dependent velocity exponent and $f \equiv (F - F_c^H)/F_c^H$ is the reduced force. Figure~\ref{fig:vinf} shows this power-law relation for $H=0.4$, $H=0.5$ and $H=0.8$, and the inset represents the exponent $\mu_H$ in terms of $H$. As is evident in this figure, $\mu_H$ increases with $H$ for the anti-correlation regime (ACR) $H\le 0.5$ implying that stronger correlations amplify the velocity growth rate once depinning occurs. It becomes more or less constant for the positive correlation regime (CR):
\begin{equation}
\begin{split}
&\left. \mu^{}_{H}\right|_{H\in \text{CR}}= 2.8\pm 0.13.
\end{split}
\end{equation}

We now examine the behavior of the model near the transition point $T=T^*$, where we observe scaling relations not previously reported in the literature. The dependence of $T^*$ on system size is smaller than our numerical errors, indicating that it is effectively insensitive to $L$. As shown in Fig.~\ref{Fig:vstar}a (shown for $H=0.4$ and $0.8$), rescaling $v^*(F)$ with $F^{-\tau_H}$ causes all curves to meet each other at the single point $T=T^*$, where $\tau_H$ is some exponent. In the vicinity of $T^*$, i.e., for $t\equiv \tfrac{T-T^*}{T^*}\ll 1$, we propose the following scaling relation:
\begin{equation}
\left. v(T,F,L)\right|_{t\ll 1}= F^{\tau_{H}}f \left( \frac{t}{F^{\gamma_H}} \right),
\label{Eq:scalingT}
\end{equation}
where $\gamma_H$ is another exponent, and $f(x)$ is a universal function which is regular at $x=0$:
\begin{equation}
f(x) =f(0) + x f'(0)+O(x^2).
\end{equation}
This relation gives rise to
\begin{equation}
F^{-\tau_{H}} v(T, F, L) = C_H + m_H(F) t+O\left(t^2\right),
\label{eq:v_scaling}
\end{equation}
where $m_H(F)\equiv \frac{B}{F^{\gamma_H}}$ is the slope in the vicinity of $t=0$, $C_H = f(0)$, $B = f'(0)$. Since $t=0$ is a crossover time, we call $\tau_H$ and $\gamma_H$ the crossover exponents, that should be determined using the data collapse analysis of the simulation data. This relation implies that at $t=0$ ($v=v^*$) we have the following scaling form:
\begin{eqnarray}
v^*(F)&=&C_{H} F^{\tau_{H}}.
\end{eqnarray}

The analysis presented in Fig.~\ref{Fig:vstar}  supports the hypothesis of Eqs.~\ref{Eq:scalingT} and~\ref{eq:v_scaling}, revealing a universal behavior in the vicinity of $T^\ast \approx 8.55\times 10^4$. Figure~\ref{Fig:vstar}(b) reports on the dependence of the scaling exponent $\tau_{H}$, and also $C_H$ on $H$, while Fig.~\ref{Fig:vstar}c presents the scaling relation between $m_H(F)$ and $F$ with the exponent $\gamma_H$ for $H=0.8$. The $H$-dependence of $\gamma_{H}$ is presented in the inset of Fig.~\ref{Fig:vstar}c. We observe that $\tau_H$ and $\gamma_H$ decrease with $H$ in the ACR $H<0.5$, while they become nearly independent of $H$ in the positive CR, $H>0.5$. More precisely, 
\begin{equation}
\begin{split}
\left. (\tau_{H},\gamma_{H})\right|_{H\in \text{CR}}= (0.155\pm 0.0001,1.0993\pm 0.003),
\end{split}
\end{equation} 
while the determination of these exponents for $H\to 0$ needs an extrapolation. Based on this observation, and as further results will confirm, this behavior persists across all the exponents examined in this work, indicating that the universal properties of the model remain robust—i.e., \textit{super-universal}—throughout the CR, while this does not hold in the ACR. \\

Figures~\ref{Fig:vc}a and~\ref{Fig:vc}b display the decay of the critical velocity $v_c(T,L)$ as a function of $T$ at $F=F_c$. While the data suggest a power-law behavior, strong fluctuations at large $T$ values limit the reliability of the fit. Therefore, the exponent $q_H$ was extracted over a more stable fitting range, spanning nearly one decade. For $H=0.4$ and $H=0.8$, the corresponding exponents are $q_H^{L=1024} = 3.895 \pm 0.1$ and $q_H^{L=1024} = 3.011 \pm 0.13$, respectively (Figs.~\ref{Fig:vc}a and~\ref{Fig:vc}b; see Eq.~\ref{Eq:Power-Law} for the definition). Figure~\ref{Fig:vc}c further illustrates the finite-size dependence of $q_H$ at $H=0.7$. From this analysis, we identify the following general behavior for all $H$ values considered in this work:
\begin{equation}
q_H^L=q_H^{L\to\infty}+\frac{A_H^q}{L},
\end{equation}
where $q_H^{L\to\infty}$ is the extrapolated exponent value in the thermodynamic limit, and $A_H^q$ is a non-universal fitting parameter, that is not important in our analysis. The corresponding thermodynamic value is presented in Fig.~\ref{Fig:vc}d, showing that $q_H^{L\to\infty}$ decreases with $H$ in the ACR, while it remains nearly constant in the CR. This trend is consistent with the behavior of $\tau_H$ and $\gamma_H$. It implies that the decay of the critical velocity is faster for smaller Hurst exponents in the ACR, whereas it becomes insensitive to variations in $H$ within the CR. The value of $q_H^{L\to\infty}$ in the CR is given by:
\begin{equation}
\begin{split}
&\left. q^{}_{H}\right|_{H\in \text{CR}}= 3.2\pm 0.2.
\end{split}
\end{equation} 
This further supports the notion of superuniversality in the CR.
\begin{figure*}[!htbp]
	\begin{subfigure}{0.45\textwidth}\includegraphics[width=\textwidth]{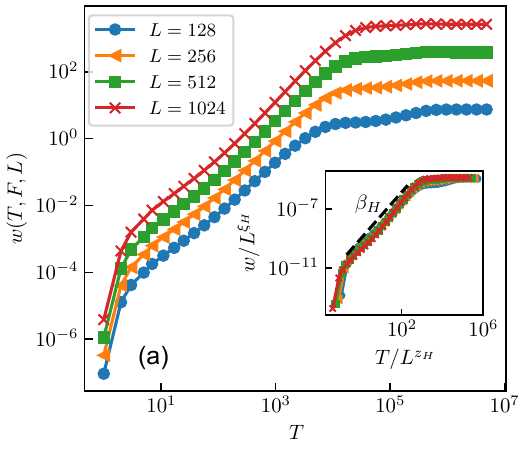}
	\end{subfigure}
	\begin{subfigure}{0.45\textwidth}\includegraphics[width=\textwidth]{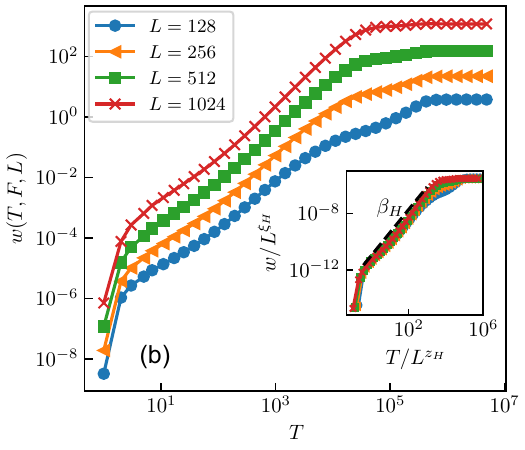}
	\end{subfigure}
    \begin{subfigure}{0.45\textwidth}\includegraphics[width=\textwidth]{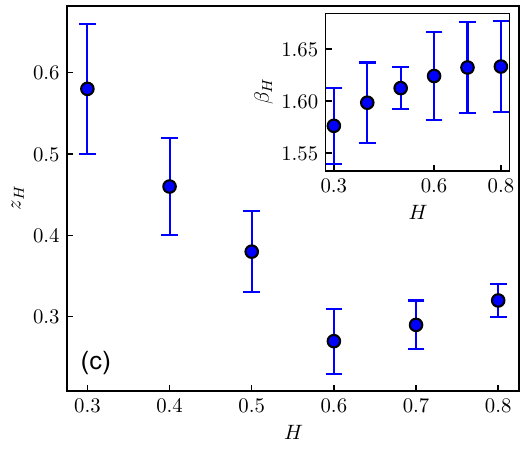}
	\end{subfigure}
     \begin{subfigure}{0.45\textwidth}\includegraphics[width=\textwidth]{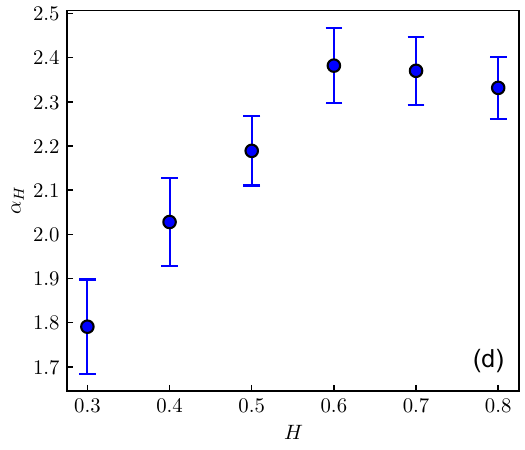}
	\end{subfigure}
	\caption{Log-Log plots of the average interface width $W$ versus time $T$ for different sizes $ L$. Insets show the data collapse analysis according to Eq.~\ref{Eq:roughnessNew}, which helps to extract $\xi_H$, $\beta_H$ and $z_H$. (a) and (b) show the results for $H=0.4$ and $H=0.8$ respectively. (c) shows $z_H$ (main) and $\beta_H$ (inset), while (d) shows $\alpha_H$ in terms of $H$.}
	\label{fig:w}
\end{figure*}

\section{Anomalous Roughness Scaling Behavior of QKPZ$_H$}\label{sec: anomalous roughness}
In Sec.~\ref{SEC:General}, we analyzed the roughness within the framework of the conventional scaling relation. Specifically, we explored the scaling behavior of this function, Eq.~\ref{Eq:roughness}, which is replaced by:
\begin{equation}
W(T,F,L)=L^{\xi_H}\mathcal{P}\left(\frac{T}{L^{z_H}}\right),
\label{Eq:roughnessNew}
\end{equation}
where $\mathcal{P}(x)$ is a universal function, and $\xi_H$ and $z_H$ denote the associated $H$-dependent scaling exponents. This behavior is supported by the data-collapse analysis shown in Figs.~\ref{fig:w}a and~\ref{fig:w}b. The reason why we call it \textit{anomalous behavior} is that, in the standard theory of rough surfaces, these exponents are connected by the hyperscaling relation $z=\xi/\beta$. In our case, however, this relation breaks down. Specifically, since $W\propto T^{\beta_H}$ at small values of $x\equiv T/L^{z_H}$, we can express $\mathcal{P}(x)=x^{\beta_H}\mathcal{G}(x)$ with $\lim_{x\to 0}\mathcal{G}(x)=\text{const}$, which yields
\begin{equation}
W(T,F,L)=T^{\beta_H} L^{\alpha_H}\mathcal{G}\left(\frac{T}{L^{z_H}}\right),
\label{Eq:w2}
\end{equation}
where 
\begin{equation}
\alpha_H \equiv \xi_H-\beta_H z_H. 
\end{equation}
In the conventional picture, hyperscaling arises from the requirement $\alpha_H\to 0$, i.e. $W$ should be independent of $L$ at very early times. As seen in Figs.~\ref{fig:w}a and~\ref{fig:w}b, this condition is not satisfied here: the early-time growth retains an additional $L$-dependence quantified by $\alpha_H$. The extracted exponents $\alpha_H$, $\beta_H$, and $z_H$ are reported in Figs.~\ref{fig:w}c and~\ref{fig:w}d. We find that, in ACR $\alpha_H$ and $\beta_H$ increase with $H$, while $z_H$ decreases with $H$. More precisely, a clear change of behavior occurs across the transition from ACR to the CR, at which the exponents remain nearly constant in accordance with $\alpha_H$, $\beta_H$, and $z_H$ . These values are:
\begin{equation}
\begin{split}
\left. (\alpha_{H},\beta_{H}, z_{H})\right|_{\text{CR}}= (&2.36\pm 0.03,1.63\pm 0.04,0.32\pm 0.03).
\end{split}
\end{equation}
The exponents are reported in Table~\ref{tab:exponents}. The set of the exponents show that QKPZ defines a different universality class in the FBM host, characterize by new scaling exponents. To quantify this more deeply, we compare it with the exponents of QEW in the FBM (QEW$_H$) background in the next section. 

\begin{table*}[!htbp]
	\caption{Numerical estimation of the exponents for the QKPZ on top of the FBM correlated lattice. }
	\begin{tabular}{| c|| c| c| c| c| c| c|}
		\hline   & $H=0.3$ & $H=0.4$ & $H=0.5$ & $H=0.6$& $H=0.7$ & $H=0.8$\\
		\hline
		\hline
		\hline  $F_{c}$  &     $1.13\pm 0.2$     & $1.0304\pm 0.02$ &   $0.9087\pm 0.14$ &     $0.8783\pm 0.15$     & $0.7783\pm 0.07$  & $0.7130\pm 0.2$ \\
		\hline $\mu_{H}$  &   $1.4556\pm 0.084$    & $2.4414\pm 0.16$  &    $2.8113\pm 0.3$ &   $2.6666\pm 0.2$    & $2.7\pm 0.275$   & $2.9414\pm 0.367$ \\
		\hline  $\tau_{H}$ &     $0.165\pm 0.0005$    & $0.16\pm 0.0005$ &   $0.156\pm 0.0003$ &     $0.155\pm 0.0005$     & $0.155\pm 0.0003$  & $0.155\pm 0.0003$ \\
		\hline $\gamma_{H}$ & $1.241\pm 0.1069$ & $1.154\pm 0.0989$ & $1.1306\pm 0.07845$ & $1.097\pm 0.08535$ & $1.104\pm 0.07735$ & $1.097\pm 0.06965$\\
		\hline $q_{H}$ & $3.984\pm 0.1$ & $3.845\pm 0.19$ & $3.33\pm 0.046$ & $3.22\pm 0.19$ & $3.15\pm 0.1$ & $3.26\pm 0.2$\\
		\hline $\alpha_H$  &     $1.791\pm 0.1$     & $2.027956\pm 0.098$ &   $2.189\pm 0.078$ &     $2.3819\pm 0.085$     & $2.3709\pm 0.077$  & $2.3316\pm 0.069$ \\
		\hline $\beta_H$  &   $1.5763\pm 0.0365$    & $1.5985\pm 0.03855$  &    $1.6125\pm 0.02$ &   $1.6241\pm 0.0419$    & $1.6323 \pm 0.04365$   & $1.6332\pm 0.044$ \\
		\hline  $z_H$ &     $0.58\pm 0.08$    & $0.46\pm 0.06$ &   $0.38\pm 0.05 $ &     $0.27 \pm 0.04$     & $0.29\pm 0.03$  & $0.32\pm 0.02$ \\
		\hline  $\xi_H$  &     $2.7\pm 0.006$     & $2.76\pm 0.007$ &   $2.8\pm 0.004$ &     $2.82\pm 0.005$     & $2.84\pm 0.003$  & $2.85\pm 0.002$ \\
		\hline
	\end{tabular}
	\label{tab:exponents}
\end{table*}
\section{Comparison of QKPZ$_H$ and QEW$_H$ universality classe}\label{sec:Comparison}
\begin{figure*}[!htbp]
	\begin{subfigure}{0.32\textwidth}\includegraphics[width=\textwidth]{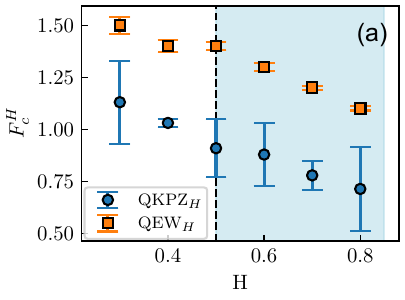}
	\end{subfigure}
	\begin{subfigure}{0.32\textwidth}\includegraphics[width=\textwidth]{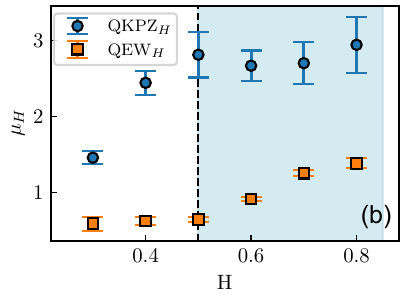}
	\end{subfigure}
    \begin{subfigure}{0.32\textwidth}\includegraphics[width=\textwidth]{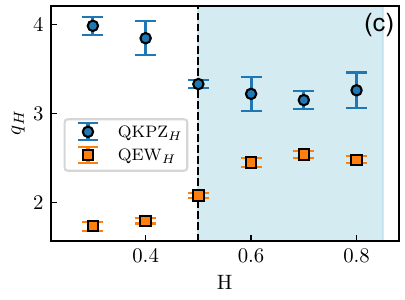}
	\end{subfigure}
     \begin{subfigure}{0.32\textwidth}\includegraphics[width=\textwidth]{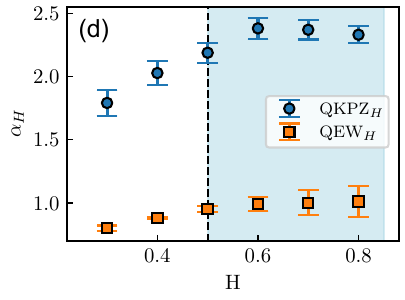}
	\end{subfigure}
    \begin{subfigure}{0.32\textwidth}\includegraphics[width=\textwidth]{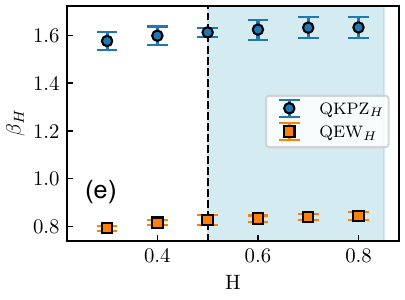}
	\end{subfigure}
    \begin{subfigure}{0.32\textwidth}\includegraphics[width=\textwidth]{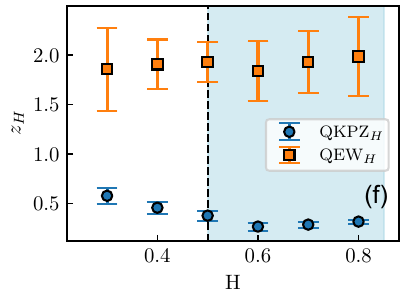}
	\end{subfigure}
    \caption{Comparison of the QEW$_H$ (after Ref.~\cite{valizadeh2023edwards}) and QKPZ$_H$ for (a) $F_c$, (b) $\mu_H$, (c) $q_H$, (d) $\alpha_H$, (e) $\beta_H$ and (f) $z_H$. The blue (right to the vertical dashed line) is CR, and the left side is ACR.}
    \label{Fig:com}
\end{figure*}

In~\cite{valizadeh2023edwards} the depinning transition was considered for QEW in the disordered FBM supports (QEW$_H$). In this subsection we provide a direct comparison between the two models, i.e. QKPZ$_H$ and QEW$_H$. The exponents are integrated in Fig.~\ref{Fig:com}. For both QEW$_H$ and QKPZ$_H$ the critical force $F_{c}$ decreases with $H$ as expected (Fig.~\ref{Fig:com}a). The absolute amounts should not be compared since they follow different normalization schemes.

While the velocity exponent $\mu_{H}$ for both models show an increase with $H$ (Fig.~\ref{Fig:com}b), we see the $\mu_H$ for the QKPZ$_H$ is meaningfully larger than the ones for the QEW$_H$. In the ACR, this exponent grows rapidly with $H$, while its growth in the this regime is much smoother for QKPZ$_H$. Similarly, the critical decay exponent $q_{H}$ is systematically larger in QKPZ$_H$ (Fig.~\ref{Fig:com}c). We observe that, $q_H$ monotonically decreases (increases) with $H$ in ACR for QEW$_H$ (QKPZ$_H$), signaling the fact that the roughness leads to higher decay rates which is attributed to the non-linear term in QKPZ$_H$.  

The roughness-related exponents also display clear contrasts. While both models exhibit anomalous roughness scaling in the presence of correlations, the QKPZ$_H$ model yields significantly larger values of $\alpha_H$ and $\beta_H$ compared to QEW$_H$ (Figs.~\ref{Fig:com}d–e), consistent with more irregular interface morphologies. The dynamic exponent $z_H$, however, is lower in QKPZ$_H$ (Fig.~\ref{Fig:com}f). Interestingly, besides the difference in the values, the behavior of $z_H$ in terms of $H$ is quite different for two models: in the ACR, while the exponent decreases in terms of $H$ for QKPZ$_H$, it does not decrease for QEW$_H$, and it is more or less constant. 

Overall, these observations demonstrate that the addition of the KPZ nonlinearity fundamentally reshapes the critical behavior: the exponents considerably change. The crossover from QEW$_H$ to QKPZ$_H$ is therefore not a simple perturbation but a qualitative shift in universality, controlled jointly by the Hurst exponent $H$ and the nonlinear growth term. This comparative analysis underlines the essential role of nonlinearities in determining the fate of driven interfaces in correlated disordered environments.

\section{Concluding Remarks}~\label{sec:Conclusion}

In this work, we investigated the depinning transition of driven interfaces in correlated disordered media by combining the quenched KPZ (QKPZ) equation with fractional Brownian motion (FBM) landscapes, controlled by the Hurst exponent $H$—which controls the strength and type of correlations in the underlying medium—(QKPZ$_H$). This approach enabled us to systematically probe how the Hurst exponent, and also the non-linear KPZ term reshape the critical properties of the transition. Our study demonstrates that correlations are not a small perturbation to classical depinning scenarios but rather play a decisive role in determining the universality of the dynamics.

The central result of this work is the general expression for the interface velocity given in Eq.~\ref{eq:velocity_piecewise}. Beyond describing the long-time dynamics, this formula also predicts the existence of a transition point $T^*$, around which specific scaling relations apply. We find that the critical force $F_{c}$ decreases monotonically with increasing $H$, indicating that positive correlations in the substrate facilitate the advance of the driven interface: the larger the value of $H$, the more easily the interface moves forward, and the smaller the corresponding $F_c$. Moreover, the dependence of $F_c$ on $H$ is found to be linear, as expressed in Eq.~\ref{Eq:Fc}.\\

The standard relations governing driven interfaces are also found to hold in this system, with critical exponents determined around $F_c$. An analytic expression for the average velocity near the transition point $T^*$ is proposed in Eq.~\ref{Eq:scalingT}, and the corresponding exponents are reported in Figs.~\ref{Fig:vstar}b and~\ref{Fig:vstar}c. Unlike uncorrelated landscapes, where pinning thresholds follow universal values, correlated environments display qualitatively different behavior: positively correlated disorder yields a weaker resisting force, thereby defining universal critical properties, whereas in the anticorrelated regime the resisting force is stronger, leading to exponents that depend explicitly on $H$.

The scaling of the steady-state velocity reveals additional richness. The velocity exponent $\mu_{H}$ grows significantly with $H$ in the anticorrelated regime, showing that once depinning occurs, correlations enhance the system’s response to the external force, producing faster motion compared to the uncorrelated case. In contrast, $\mu_{H}$ remains nearly constant in the correlated regime. At the depinning threshold, temporal relaxation is slower in correlated media, as indicated by the decrease of the decay exponent $q_{H}$. This behavior is natural: higher values of $H$ favor the moving phase, which delays relaxation at criticality.

The other part of the work concerns the roughness of the interface. Our results reveal anomalous scaling (\ref{Eq:w2}) beyond the standard Family–Vicsek form~\ref{Eq:roughness}. Both the roughness exponent $\alpha_{H}$ and the growth exponent $\beta_{H}$ increase with $H$ in the anticorrelation regime, and become nearly constant in the correlation regime. At the same time, the dynamic exponent $z_{H}$ decreases, implying that temporal correlations shorten even as spatial irregularities intensify. Together, these shifts point to the emergence of a new scaling structure that departs from conventional universality and is fully controlled by the degree of correlations in the disorder. We see that the non-linearity term in QKPZ$_H$ changes considerably the universality class of QEW$_H$. It is also very different from the conventional universality classes present in the literature. The exponents are reported in Table~\ref{tab:exponents}.\\

In summary, our findings establish that the depinning transition of QKPZ$_H$ interfaces in correlated media is characterized by continuously varying critical exponents, tunable through the Hurst exponent $H$. This challenges the notion of a single universality class for depinning and demonstrates instead a spectrum of critical regimes shaped by the correlation properties of the medium, as well as the non-linearity in the growth dynamics. Such results underscore the fundamental importance of disorder correlations in non-equilibrium interface dynamics and provide a clear framework for interpreting deviations from classical universality observed in both simulations and experiments.

\appendix
\section{Scaling Arguments}\label{App:Scaling}

This appendix details the scaling properties of Fractional Brownian Motion (FBM). The process is defined by its correlation function (Eq.~\ref{Eq:def}):
\begin{equation}
	\langle [B_{H}(\mathbf{r})-B_{H}(\mathbf{r}')]^{2}\rangle \propto |\mathbf{r}-\mathbf{r}'|^{2H}.
 	\label{Eq:defA}
\end{equation}

To analyze its scaling behavior, we apply a transformation $\mathbf{r}\rightarrow \alpha \mathbf{r}$, where $\alpha > 0$ is a scaling factor. For Eq.~\ref{Eq:defA} to remain consistent under this transformation, the FBM field must satisfy the following scaling relation:
\begin{equation}
	B_H(\alpha\mathbf{r}) \stackrel{d}{=} \alpha^{H} B_H(\mathbf{r}),
	\label{A1}
\end{equation}
where $\stackrel{d}{=}$ denotes equality in probability distributions.

Our objective is to demonstrate that the FBM generation method described in the main text produces a field that adheres to this property. The method begins by generating an uncorrelated Gaussian white noise field $\zeta(\mathbf{r})$ across a lattice, characterized by:
\begin{equation}
	 \langle \zeta(\mathbf{r}) \rangle = 0, \quad \langle \zeta(\mathbf{r})\zeta(\mathbf{r}') \rangle = \delta^{(2)}(\mathbf{r}-\mathbf{r}').
	\label{Eq:zetaA}
\end{equation}
The scaling property of the Dirac delta function,
\begin{equation}
	\langle \zeta(\alpha\mathbf{r})\zeta(\alpha\mathbf{r}') \rangle = \alpha^{-2} \delta^{(2)}(\mathbf{r}-\mathbf{r}'),
	\label{Eq:zetaA2}
\end{equation}
implies that the noise itself scales as:
\begin{equation}
	\zeta(\alpha \mathbf{r}) \stackrel{d}{=} \alpha^{-1} \zeta(\mathbf{r}).
	\label{Eq:zetaScaling}
\end{equation}

We now examine the scaling behavior in Fourier space. The Fourier transform of the noise is defined as:
\begin{equation}
\tilde{\zeta}(\mathbf{k}) \equiv \frac{1}{L} \sum_{\mathbf{r}} \zeta(\mathbf{r}) e^{-i\mathbf{k} \cdot \mathbf{r}},
\end{equation}
where $L = \sqrt{L_x L_y}$ for a square lattice ($L_x = L_y$). Applying the scaling relation from Eq.~\ref{Eq:zetaScaling} yields:
\begin{equation}
	\frac{1}{L} \sum_{\mathbf{k}} \tilde{\zeta}(\mathbf{k}) e^{i\alpha\mathbf{k} \cdot \mathbf{r}} = \alpha^{-1} \frac{1}{L} \sum_{\mathbf{k}} \tilde{\zeta}(\mathbf{k}) e^{i\mathbf{k} \cdot \mathbf{r}}.
\end{equation}

Transitioning to the thermodynamic limit, where $\sum_{\mathbf{k}} \rightarrow \left( \frac{L}{2\pi} \right)^2 \int \mathrm{d}^2\mathbf{k}$, this equation becomes:
\begin{equation}
	\frac{1}{\tilde{L}} \left( \frac{\tilde{L}}{2\pi} \right)^2 \int \mathrm{d}^2\mathbf{Q} \, \tilde{\zeta} \left( \frac{\mathbf{Q}}{\alpha} \right) e^{i\mathbf{Q} \cdot \mathbf{r}} = \frac{1}{L} \left( \frac{L}{2\pi} \right)^2 \int \mathrm{d}^2\mathbf{k} \, \tilde{\zeta}(\mathbf{k}) e^{i\mathbf{k} \cdot \mathbf{r}},
\end{equation}
where we have made the change of variables $\mathbf{Q} \equiv \alpha \mathbf{k}$ and defined a scaled system size $\tilde{L} \equiv \alpha^{-1} L$. This leads to the key scaling relation for the Fourier-transformed noise:
\begin{equation}
	\tilde{\zeta}(\alpha^{-1} \mathbf{k}) \stackrel{d}{=} \tilde{\zeta}(\mathbf{k}).
	\label{Eq:FourierNoiseScaling}
\end{equation}
(Note: While the specific pre-factors in the Fourier transform definition can affect the exact form of this relation, the final result for $B_H$ remains unchanged.)

The FBM field in Fourier space is constructed by filtering the noise:
\begin{equation}
\tilde{B}_H(\mathbf{k}) = \mathbf{k}^{-H-1} \tilde{\zeta}(\mathbf{k}),
\end{equation}
as given in Eq.~\ref{Eq:fBM}. Using the scaling relation for the noise (Eq.~\ref{Eq:FourierNoiseScaling}), we find the corresponding scaling for $\tilde{B}_H$:
\begin{align}
\tilde{B}_H(\alpha^{-1}\mathbf{Q}) &= (\alpha^{-1}\mathbf{Q})^{-H-1} \tilde{\zeta}(\alpha^{-1}\mathbf{Q}) \\\nonumber
&= \alpha^{H+1} \mathbf{Q}^{-H-1} \tilde{\zeta}(\mathbf{Q}) \\\nonumber
&= \alpha^{H+1} \tilde{B}_H(\mathbf{Q}).
\label{Eq:FBMFourierScaling}
\end{align}

Finally, we confirm that this leads to the desired real-space scaling by performing the inverse Fourier transform:
\begin{widetext}
\begin{align}
B_H(\alpha\mathbf{r}) &= \frac{1}{L} \left( \frac{L}{2\pi} \right)^2 \int \mathrm{d}^{2}\mathbf{k} \, \tilde{B}_H(\mathbf{k}) e^{i\alpha\mathbf{k} \cdot \mathbf{r}} \\ \nonumber
&= \frac{\alpha^{-1}}{\tilde{L}} \left( \frac{\tilde{L}}{2\pi} \right)^2 \int \mathrm{d}^{2}\mathbf{Q} \, \tilde{B}_H(\alpha^{-1}\mathbf{Q}) e^{i\mathbf{Q} \cdot \mathbf{r}} && \text{(substitute } \mathbf{Q} = \alpha\mathbf{k}) \\ \nonumber
&= \frac{\alpha^{-1}}{\tilde{L}} \left( \frac{\tilde{L}}{2\pi} \right)^2 \int \mathrm{d}^{2}\mathbf{Q} \, \left[ \alpha^{H+1} \tilde{B}_H(\mathbf{Q}) \right] e^{i\mathbf{Q} \cdot \mathbf{r}} && \text{(apply Eq.~\ref{Eq:FBMFourierScaling})} \\ \nonumber
&= \alpha^{H} \left[ \frac{1}{\tilde{L}} \left( \frac{\tilde{L}}{2\pi} \right)^2 \int \mathrm{d}^{2}\mathbf{Q} \, \tilde{B}_H(\mathbf{Q}) e^{i\mathbf{Q} \cdot \mathbf{r}} \right] \\ \nonumber
&= \alpha^{H} B_H(\mathbf{r}).
\end{align}
\end{widetext}
This result confirms that the generated FBM field indeed obeys the scaling law stated in Eq.~\ref{A1}.
\section*{Data availability}
The findings of this study can be supported by data that is accessible from the corresponding author upon making a reasonable request.
\section*{Author contributions}
N.V. wrote the code and did the simulations, prepared the data and figures, and wrote the first version of the manuscript. M.N.N. designed the problem, Supervised the work analyzed the data and wrote the final version of the paper.
\section*{Declaration of competing interest}
The authors declare no competing interests.
\section*{Acknowledgments}
The authors gratefully acknowledge the Iranian National Science Foundation (INSF), Iran (Project No.4036773), for financial support. We also thank the Department of Physics at the University of Mohaghegh Ardabili for providing computational facilities and a stimulating research environment. 
\newpage
\bibliography{refs}
%\bibliography{MyReferences}
\end{document}